\documentclass[a4paper]{article}
\usepackage[a4paper,
            bindingoffset=0.2in,
            left=1in,
            right=1in,
            top=1in,
            bottom=1in,
            footskip=.25in]{geometry}
\usepackage{authblk}
\usepackage{blindtext}
\usepackage[square,numbers]{natbib}
\title{Black hole models and astronomical observations}
\author[1]{Serge Parnovsky}
 \affil[1]{Taras Shevchenko National University of Kyiv, Astronomical observatory, Observatorna str. 3, Kyiv 04053, Ukraine; parnovsky@knu.ua}
\begin{document}
\maketitle
\begin{abstract}
Some well-known metrics used as models of static non-rotating black holes are considered. These are the Reissner-Nordstr{\"o}m-de Sitter, Bardeen-Kiselev and McVittie models. A significant difference between them in how adequately they can describe or be the basis for describing observed astronomical objects is demonstrated. A simple method for constructing new models of static non-rotating black holes is presented. It has been shown that a region of gravitational repulsion may exist around the horizon of a Bardeen-Kiselev black hole.


\end{abstract}

\section{Introduction}
Astronomers have discovered a large number of massive compact objects that look like black holes (BHs) and may turn out to be exactly BHs. They attract surrounding bodies. The accretion of matter is accompanied by processes that lead to the generation of radiation observed on Earth. The mass of compact objects varies widely from stellar ones of several solar masses to galactic ones (supermassive BHs with a mass of up to $10^{11}$ solar masses). The Nobel Prize in Physics in 2020 was given for the discovery of a supermassive compact object at the centre of our galaxy. 

BHs are particularly valuable objects for study by physicists. Therefore, it is desirable to have an adequate model of black holes. Over the years of the existence of GTR, dozens of models have been proposed, but their value varies greatly.

When we come across a scientist's name next to the name of an object in a scientific article, we are usually talking about a particular case of this object. For example, Seyfert or Markaryan galaxies are particular types of galaxies. Comets Galey or Emcke are comets named after their discoverers. It is possible to deal with different models to describing the same object. Examples include Majorana and Dirac neutrinos.
 
But the situation is significantly more complicated when it comes to BHs. Often, adding a name implicitly indicates that we are not talking about an object observed by astronomers, but about a hypothetical or speculative object. Its explanation is based on assumptions, among which there is at least one of the following statements:
\begin{itemize}
\item Gravity is described not by the General Relativity (GR), but by another theory;
\item Electromagnetic interaction is described by an alternative theory;
\item The black hole has a magnetic charge;
\item The presence of dark energy (DE) or a matter with very specific properties is necessary.
\end{itemize}

Sometimes the object under study does not belong to BHs at all, but is a naked singularity (NS). Complex composite objects like gravastars \cite{p1} with DE-filled cavities surrounded by shells are sometimes considered as possible BH models.

A specialist in the BH issue can understand what a particular article is about. However, a scientist whose area of scientific interest lies elsewhere may get a false impression about the status of the object being studied. Often in articles using obviously inadequate models, the authors do not mention the shortcomings of the model. The title of the article sometimes mentions BHs without adding any model name, and the introduction does not indicate the non-standard approach. Moreover, it usually mentions the observed astronomical objects, without adding that they are unlikely to be described by the specific model used in the article. The conclusions often refer not to considered BH model, but to BHs in general.
 
This paper is related to methodological aspects of BHs research that are not usually discussed in scientific articles. However, some of the articles I have read in or reviewed for scientific journals make me confident that such an article could help the reader whose field of research is not directly related to the study of BHs. Let’s discuss some of the nuances that are worth knowing to properly understand and classify articles in this field of science.

To paraphrase the quote from the satirical allegorical novella ``Animal Farm'' by George Orwell, we can say that ``All black hole models are equal, but some models are more equal than others.'' I demonstrate this with two metrics, namely the Bardeen-Kiselev \cite{p2} and the McVittie \cite{p3} ones. 
These solutions are compared with the standard Reissner-Nordstr{\"o}m-de Sitter (RNdS) BHs. All of them describe the simplest case of static non-rotating BHs.

\section{Non-rotating spherically symmetrical black holes}\label{S1}

Let us consider the solutions of Einstein’s equations in their classical form
\begin{equation}
\label{a1}
R_{ik}=\kappa \left(T_{ik}-\frac{1}{2}T\right),\quad \kappa=\frac{8\pi G}{c^4}.
\end{equation}
I use the same notation and signs as the well-known textbook \cite{p4} and a system of units in which the speed of light $c$ and the gravitational constant $G$ are equal to unity. Latin indices take values 0,1,2,3. DE is included in the energy-momentum tensor $T_{ik}$.

Let's start with the static space-time metric, i.e. non-rotating BH, having the form
\begin{equation}
\label{a2}
ds^2=e^{\nu(r)}dt^2-e^{\lambda(r)}dr^2-r^2\left(d\Theta^2+\sin^2\Theta d\varphi^2\right).
\end{equation}
If the energy-momentum tensor satisfies the condition
\begin{equation}
\label{a3}
T_0^0=T_1^1,
\end{equation}
then metric (\ref{a2}) can be transformed after change the time variable to the form (see Section 97 in \cite{p4})
\begin{equation}
\label{a4}
ds^2=f(r)dt^2-\frac{1}{f(r)}dr^2-r^2\left(d\Theta^2+\sin^2\Theta d\varphi^2\right).
\end{equation}

\subsection{Standard Reissner-Nordstr{\"o}m-de Sitter black holes}

Condition (\ref{a3}) is satisfied by energy-momentum tensors of the DE in the form of the cosmological constant $\Lambda$, the centrally symmetric radial electric field in classical electrodynamics, and vacuum. The metric of a static black hole with mass $M$ and electric charge $Q$ has the form (\ref{a4}) with 
\begin{equation}
\label{a5}
f=1-\frac{2M}{r}+\frac{Q^2}{r^2}-\frac{\Lambda}{3}r^2.
\end{equation}
This solution is usually called RNdS one. The central singularity corresponds to $r=0$. The horizons correspond to the values of the coordinate $r$ at which the function $f(r)$ becomes zero.

The metric (\ref{a4},\ref{a5}) is an exact solution of the Einstein equations (\ref{a1}) with the corresponding energy-momentum tensor for any values of the constants $M$, $Q$ and $\Lambda$. These quantities are independent from each other. Special cases of this solution are named after their discoverers.

For $\Lambda=0$ this is the Reissner-Nordstr{\"o}m solution obtained in \cite{p5} and \cite{p6}. If $M\geq|Q|$, then it has at least one event horizon and we are talking about a Reissner-Nordstr{\"o}m black hole. For $M>|Q|$ there are two concentric event horizons. They become degenerate for $M=|Q|$, which corresponds to an extreme black hole. In this case, the two roots of the equation $f(r)=0$ coincide. For $M<|Q|$ there is no horizon and space-time contains naked singularity (NS).

For $Q=\Lambda=0$ this is the Schwarzschild solution \cite{p7}. If $M>0$, it describes an uncharged, non-rotating Schwarzschild BH, if $M<0$, then it is a NS. The geodesic structure of BHs (\ref{a4},\ref{a5}) at $\Lambda=0$ is described in \cite{p8}. For Schwarzschild BH it is significantly different from Reissner-Nordstr{\"o}m BH, not to mention space-time (\ref{a4},\ref{a5}) with NS. The solution for $Q=0, \Lambda > 0$ was obtained by Kottler \cite{p9}. This case is usually called a Schwarzschild-de Sitter black hole.

So, the metric (\ref{a4},\ref{a5}) is an exact solution of the Einstein-Maxwell equations for any parameters included in it. For certain relationships between the parameters it describes BHs with one to three horizons. One event horizon is present at $\Lambda=0$ in the Schwarzschild BH ($Q=0$, $m>0$) and the degenerate Reissner-Nordstr{\"o}m BH ($Q=m$). Two event horizons are present in the Reissner-Nordstr{\"o}m BH ($Q>m$, $\Lambda=0$). One event horizon and one cosmological horizon are present in the Kottler (Schwarzschild-de Sitter) BH ($Q=0$, $m>0$, $\Lambda>0$).

I mentioned the possibility of multiple roots of the equation $f(r)=0$. Theoretically, one can consider the speculative case of the coincidence of all its roots. This requires an unambiguous relation between all three parameters: $\Lambda=r_g^{-1}$, $M=r_g/2$, $Q=3^{-1/2}r_g$. Here $r_g$ is a triple root value. The space-time described by the metric (\ref{a4},\ref{a5}) in this case has unusual properties. However, this exotic solution obviously does not describe the observed astronomical objects.

Solution (\ref{a4},\ref{a5}) allows only matter and fields with an energy-momentum tensor satisfying condition (\ref{a3}). Neither baryon nor dark matter satisfies it, so their presence is formally incompatible with the RNdS metric. Therefore, it is used to describe the spacetime metric near a BH as an approximate solution, and consider the motion of matter and photons against this background. As a result of the infall of matter, the mass of the black hole increases, but slowly enough, so it can be considered constant when constructing accretion models or modeling the observed accretion disks around supermassive BHs.

Astronomers observe many compact massive objects that could be BHs. If they do not rotate, then the metric (\ref{a4},\ref{a5}) is a good model for their approximate description. The model is approximate one because the object is not static, it is surrounded by an accretion disk and other infalling matter. If it rotates, then the model is based on the Kerr metric.

\section{What makes us use other BH models?}

Why do scientists not limit themselves to the RNdS metric when describing black holes? There are at least three reasons for this, from objective to subjective. They often complement each other.

The first reason is related to the fact that the Universe clearly contains matter, both baryonic and dark. Condition (\ref{a3}) is not satisfied for it, so the solution (\ref{a4},\ref{a5}) cannot describe a BH surrounded by matter. Far from the BH, we can consider a homogeneous isotropic space-time described by the Friedmann-Lema\^{i}tre-Robertson-Walker (FLRW) solution \cite{p4}, but we do not know the exact solution describing a BH against the FLRW background. This is related both to the complexity of the GTR equations and to the fact that by placing a black hole in a homogeneous distribution of matter, we immediately violate both the homogeneity of space and the BH stationarity.

Indeed, matter under the influence of BH gravity begins to move towards it, creating a density gradient. It falls in BH, increasing the BH mass. Real astronomical objects do not have spherical symmetry, which further complicates the consideration of the processes occurring near it. Without knowing the exact solution in the presence of matter, scientists are looking for at least approximate metrics to describe space-time both near and far from the BH in the case of spherical symmetry. More than one such metric has been proposed, but we should not forget that these solutions are not exact ones.

The second reason is related to Einstein's equations. Although GR has successfully passed all tests so far, we cannot be completely sure that it will not be replaced by another theory of gravity. There are many alternative theories within which one can consider the properties of a black hole, if they exist in a specific theory. In addition, one can use one of its alternatives instead of classical electrodynamics and choose any of the numerous models to describe DE. Using alternative theories of gravity and/or electromagnetism can provide new BH model and obviously gives the opportunity to publish an article with results obtained for the first time.

The third reason is related to the presence of the central singularity with infinitely large curvature and tidal forces at $r=0$ in the solution (\ref{a4},\ref{a5}). There is also a singularity in the Kerr metric. Moreover, according to the theorem proved by Penrose in 1965, within the framework of general relativity, inside each black hole some kind of singularity exists, provided that some fairly reasonable assumptions are met \cite{p10}. This result was not liked by some experts in this field, hoping to obtain a solution with a black hole free of singularity. Such hypothetical objects are called regular black holes. A number of papers have attempted to get a regular BH metric by taking into account quantum effects, modifying the theory of gravity, or considering media in which the restrictions used in Penrose's theorem did not hold. 

\section{An easy way to create a new BH model}

Among other approaches, one can use the simplest one. A scientist choose the desired type of metric, say (\ref{a4}), and propose a function $f(r)$ that satisfies his or her wishes. It should be defined at $r\geq 0$ and have no discontinuities. If neither $f$ nor its derivative go to infinity at $r \geq 0$ we get a regular BH. If $f>0$, then the coordinate $r$ is spatial, and $t$ is temporal. Regions with $f<0$ are inside the BH or outside the cosmological horizont, in them the coordinate $t$ is spatial, and $r$ corresponds to time. The values of the coordinate $r$, at which the function $f$ changes sign, correspond to event horizons. Having chosen an appropriate function $f$, which is negative at $r<r_0$, positive at $r>r_0$, with $f\to 1$ at $r\to \infty$, and vanishing at $r=r_0$, we obtain a metric of the form (\ref{a4}), which can pretend to be a BH model in an asymptotically flat universe. Substituting this metric into Einstein's equations (\ref{a1}) one can calculate the energy-momentum tensor $T_{ik}$ required for this invented BH model to correspond to GTR. 
\begin{equation}
\label{at}
\kappa T_0^0=\kappa T_1^1=\frac{1-f}{r^2}-\frac{f^\prime}{r},\quad \kappa T_2^2=\kappa T_3^3=-\frac{f''}{2}-\frac{f^\prime}{r}.
\end{equation}
The prime denotes the derivative with respect to $r$. Naturally, this tensor satisfies condition (\ref{a3}), but for regular BH it does not satisfy the conditions assumed in the proof of Penrose's theorem.

The final stage consists of attempts to somehow explain the form of the obtained energy-momentum tensor. Everything is used: dark matter and dark energy with exotic equations of state, magnetic monopoles, etc. If condition (\ref{a3}) is not satisfied for some components, then it is declared that they are present in such a proportion that this condition is satisfied for the full tensor $T_{ik}$. If it is not possible to somehow explain the form of the energy-momentum tensor, then changing the expression for the action of gravitational or electromagnetic fields can help, i.e., switching to alternative theories of these fields.

If this does not solve the problem, then it remains to write that the necessary energy-momentum tensor is provided by undiscovered fields or quantum effects.

\section{Bardeen black holes}

Let us consider a number of specific models of black holes, starting from the Bardeen BH. Bardeen proposed one of the first models of regular BH \cite{p11} in 1968 when Penrose's theorem had not yet received proper assessment from the gravitational community and singularity-free solutions were actively being sought. It has a form (\ref{a4}) with
\begin{equation}
\label{a6}
f=1-\frac{2m(r)}{r},\quad m(r)=M\left[ 1+\left(\frac{q}{r}\right)^2\right]^{-3/2}.
\end{equation}
The original proceedings are not readily available outside the former USSR, but can be found in the libraries listed at https://hsm.stackexchange.com/questions/14355/james-m-bardeens-missing-talk. Bardeen black hole (\ref{a4},\ref{a6}) is a spherically symmetric solution without cosmological constant. In 1968, before the discovery of the accelerated expansion of the Universe, $\Lambda$ was considered as an optional element of GRT. This component was added to the model in the article \cite{p12}.The metric has the form (\ref{a4}) with
\begin{equation}
\label{a7}
f=1-\frac{2m(r)}{r}-\frac{\Lambda}{3}r^2.
\end{equation}
Metrics (\ref{a6}) and (\ref{a7}) far from the BH become the Schwarzschild and Schwarzschild-de Sitter solutions.

What do these solutions correspond to? According to Ayon-Beato and Garcia, they describe the gravitational field of a nonlinear magnetic monopole, i.e., a solution to Einstein equations coupled to a nonlinear electrodynamics \cite{p13}. The letter $q$ denotes the magnetic charge of the monopole. The mass function that generated the Bardeen metric was obtained in \cite{p14}. A completely different explanation of Bardeen regular black hole as a quantum-corrected Schwarzschild black hole was proposed in \cite{p15}.

\section{Bardeen-Kiselev black holes}

The Bardeen-Kiselev BH metric was proposed in the paper \cite{p2}. It has the form (\ref{a4}) with
\begin{equation}
\label{a8}
f=1-\frac{2m(r)}{r}-\frac{2C}{r^{3w+1}}.
\end{equation}
It has two additional constants $C > 0$ and $-1 \le w \le 0$ compared to (\ref{a4},\ref{a7}). They describe an environment surrounding the BH, which Kiselev calls quintessence. The pressure and density of the quintessence are usually related by the equation of state
\begin{equation}
\label{a9}
p_q=w\rho_q.
\end{equation}
Its density depends on the radial coordinate as
\begin{equation}
\label{a10}
\rho_q=-\frac{3wC}{2}r^{-w-3}.
\end{equation}
This quintessence does not fall on the BH, the mass of the latter does not change and the metric is constant in time.

A term with a cosmological constant was added to the function (\ref{a8}) in the paper \cite{p16}:
\begin{equation}
\label{a11}
f=1-\frac{2m(r)}{r}-\frac{\Lambda}{3}r^2-\frac{2C}{r^{3w+1}},\quad m(r)=M\left[ 1+\left(\frac{q}{r}\right)^2\right]^{-3/2}.
\end{equation}
Note that the term quintessence is usually used to denote a special kind of DE. However, in the spacetime described by the metric (\ref{a4},\ref{a11}), there is also a DE in the form of the cosmological constant $\Lambda$, so that there are two different kinds of DE in the model. However, the interpretation of the medium surrounding the Bardeen-Kiselev BH has changed over time. Visser showed in the article “The Kiselev black hole is neither perfect fluid, nor is it quintessence” that its energy-momentum tensor is anisotropic \cite{p17}. The equation of state (\ref{a9}) refers to the average pressure equal to one-third of the sum of the radial and twice the tangential pressure. So the medium is not simply anisotropic, the axis of its anisotropy at any point in space is directed toward the Kiselev BH. However, equation (\ref{at}) shows that this is the only possible type of anisotropy.

Let us consider in more detail how the solution (\ref{a4},\ref{a11}) was obtained in the article \cite{p16} within the framework of GTR with a cosmological constant $\Lambda$ and in the presence of a medium that is a combination of fluid and a magnetic field described by unusual nonlinear electrodynamics. In the coordinates of the metric (\ref{a4}), the only nonzero component of the Maxwell-Faraday tensor $F^{ik}$ is the radial magnetic field with $F_{23}=-F_{32}=q\sin\Theta$, where $q$ is the magnetic charge of Kiselev BH. The field invariant in this case is the scalar $F=F^{ik}F_{ik}=2q^2r^{-4}$. Another scalar invariant $e_{iklm}F^{ik}F^{lm}$ is zero.

The action is described by the standard expression
\begin{equation}
\label{a12}
S=\int d^4x\sqrt{-g}(R+2\Lambda+L).
\end{equation}
The Lagrangian $L$ describes the contribution of the anisotropic fluid and electromagnetic field in the framework of nonlinear electrodynamics and depends on the parameters $F$, $q$, $w$ and $C$. It defines the energy-momentum tensor of matter and the magnetic field. Since its form is obtained from equations (\ref{a1}) with the metric (\ref{a4},\ref{a11}), one can determine the form of the Lagrangian necessary for the existence of Bardeen-Kiselev BHs.
\begin{equation}
\label{a13}
L=\frac{3\sqrt{2}M}{\pi\sqrt{q}\left(\sqrt{\frac{2}{F}}+2q\right)^{5/2}}-\frac{3}{4\pi}wC\left(\frac{2F}{q^2}\right)^{3(w+1)/4}.
\end{equation}
At $C=0$, i.e. in the absence of fluid, the second term vanishes and we obtain the Lagrangian for Bardeen BH nonlinear electrodynamics. It is evident that it does not transform into the Lagrangian of classical electrodynamics for a weak field. Since the correctness of the description of weak electromagnetic fields by classical electrodynamics has been repeatedly confirmed experimentally, the obtained Lagrangian does not describe the world around us. Moreover, in the limit $q\to 0$ it diverges as $q^{-3}$.
 
In addition, the expression clearly includes the parameter $M$, corresponding to the mass of the BH in the Bardeen model. It is impossible to assume that the laws of electrodynamics would depend on the mass of any cosmic object. The situation is especially nontrivial when there are at least two Bardeen BHs with different values of $M$ in the universe. Naturally, the expression for the field Lagrangian should not include the value of the magnetic charge $q$ of a particular BH.

Let us turn to the second term in (\ref{a13}). It should describe the contribution of the anisotropic fluid. However, it clearly includes both the electromagnetic field invariant $F$ and the magnetic charge $q$. In fact, it is clear that they are included in the combination $2Fq^{-2}=4 r^{-4}$. This term in reality depends not on the field invariant, but on the distance $r$ to the BH. In other words, the properties of the medium have to change depending on the distance from the BH.
 
It is clear from (\ref{a13}) that the medium and the field around the Bardeen and Bardeen-Kiselev BHs must have properties completely different from those observed in nature. Therefore, the observed astronomical objects cannot be described by these solutions.

Let us formulate conclusions about Bardeen and Bardeen-Kiselev BHs
\begin{enumerate}
\item These solutions describe a static space-time with spherical symmetry, which at large r goes over to the Minkowski metric at $\Lambda=0$ or to the de Sitter metric at $\Lambda \neq 0$.
\item It has an event horizon at the coordinate value $r$, which is the root of the equation $f(r)=0$.
\item To implement the solution, the presence of a medium with a specific energy-momentum tensor is necessary.
\item There are different possible versions of what exactly can provide the required energy-momentum tensor. They include a declaration of the existence of fields unknown to science or the action of quantum effects.
\item The most common interpretation, remaining after discarding particularly exotic explanations, suggests two new factors for Bardeen BH. This is the special type of nonlinear electrodynamics to describe the electromagnetic field and the existence of a magnetic charge of the black hole.
\item The electromagnetic field Lagrangian needed to provide the required energy-momentum tensor has nothing in common with the classical electrodynamics Lagrangian and does not transform into the latter in any limit. A field with such a Lagrangian is described by equations different from Maxwell's equations, which have been repeatedly confirmed experimentally. The formula for the Lagrangian (\ref{a13}) contains the parameters of a specific BH and, therefore, cannot be universal.
\item Bardeen-Kiselev BH additionally contains an anisotropic fluid with a Lagrangian that directly includes the coordinate $r$, although in a slightly disguised form. Thus, the properties of this fluid change depending on the distance from the BH at the same density, which is hardly possible in nature.
\end{enumerate}

\section{Zones of gravitational attraction and repulsion around\\ Bardeen-Kiselev BH}\label{S2}

Let us consider separately the question of matter falling onto a BH. All BHs discovered by astronomers are observed because they attract surrounding bodies. They either orbit the BH or fall onto it. BHs are surrounded by accretion disks, in which matter orbits around and then falls. The fall is accompanied by the generation of electromagnetic radiation in a wide frequency spectrum, which reaches the Earth. The Bardeen-Kiselev BH models considered above do not include ordinary matter. But we can consider its interaction with the BH as the motion of a test body in the BH gravitational field. If the particle is initially motionless and the spatial components of its four-dimensional velocity $u^i$ are zero, then the magnetic field does not act even on charged particles if their magnetic moment is zero. So the interaction force is purely gravitational. The question may arise as to how applicable these considerations are in the case of electrodynamics with a non-standard Lagrangian (\ref{a13}). This changes the form of the second pair of Maxwell's equations. The equation of motion of a charged particle in an electromagnetic field does not change, as can be seen from their derivation in \cite{p4}. Therefore, the magnetic field in the Bardeen-Kiselev model also does not act on a stationary particle without a magnetic moment.

The change in the radial component of the velocity is obtained from the equation of motion \cite{p4}
\begin{equation}
\label{a14}
\frac{d^2r}{ds^2}=-\Gamma^1_{ik}u^iu^k.
\end{equation}
Here $\Gamma^l_{ik}$ are the Christoffel symbols. For the metric (\ref{a4}) and the stationary particle with $u^i=(f^{-1/2},0,0,0)$ we obtain
\begin{equation}
\label{a15}
\frac{d^2r}{ds^2}=\frac{f^\prime}{2f}.
\end{equation}
The sign of acceleration is determined by the sign of the derivative $df/dr$. Attraction or repulsion corresponds to a decrease or increase in the velocity of removal from the center with time during radial motion. Let us consider the simplest case of Kiselev BH without the cosmological constant (\ref{a6}) outside its horizon, i.e. for $f>0$. It is easy to verify that attraction to the BH is observed for $r>r_0$, where $r_0=\sqrt{2} q$ is denoted. Since $f(r_0)=1-4M/(3^{3/2}q)$, then for $4M<3^{3/2}q$ there have to be a region between $r_0$ and the horizon in which the BH repels stationary test particles. For an especially large magnetic charge, repulsion does not allow particles to fall inside the BH even after acceleration in the region of attraction. This is clearly different from the properties of objects observed by astronomers.

Considering more complex Bardeen-Kiselev BHs, one can ask what keeps the fluid from falling onto the black hole, especially if the horizon lies in the region of attraction. As it falls, the metric ceases to be static, and the mass M increases. Note that in flat space, the mass of matter with density varying according to law (\ref{a10}) and located inside a sphere of radius $R$ is finite, and outside the sphere it is infinite. Therefore, one can assume that even taking into account the curvature of space, the fluid is not concentrated mainly around the BH, but fills the entire universe up to the cosmological horizon. However, its density decreases quickly enough that the fall onto the BH could compensate for the loss of mass of the fluid falling through the horizon onto the BH. Note that even if the BH is surrounded by a region with gravitational repulsion, the density and pressure of the fluid increase as it approaches the BH, which contradicts the law of hydrostatic equilibrium.

\section{McVittie black holes}

In 1933 McVittie proposed a metric that would describe a BH embedded in FLRW spacetime \cite{p3}. This solution is still used today. In general, the metric in isotropic coordinates has the form 
\begin{equation}
\label{a16}
ds^2=\left(\frac{1-\mu(t,\tilde{r})}{1+\mu(t,\tilde{r})}\right)^2dt^2-\frac{(1+\mu(t,\tilde{r}))^4}{K^2(\tilde{r})}a^2(t)\left[d\tilde{r}^2+\tilde{r}^2\left(d\Theta^2+\sin^2\Theta d\varphi^2\right)\right],
\end{equation}
\begin{equation}
\label{a16a}
K(\tilde{r})=1+\frac{k\tilde{r}^2}{4},\quad \mu(t,\tilde{r})=\frac{mK^{1/2}(\tilde{r})}{2\tilde{r}a(t)}.
\end{equation}
The notations used are: $a(t)$ is the scale factor of the Universe as a function of the cosmic time $t$, $k$ is the sign of the spatial curvature (flat universe $k = 0$, closed universe $k = 1$, open universe $k = -1$), and $m$ is a constant parameter related to the mass of the BH. At $m=0$ (\ref{a16},\ref{a16a}) goes transformed into the FLRW metric.
 
If we are interested in the region near the BH with a mass significantly smaller than the radius of curvature of the Universe, then we can consider an asymptotically flat model with $k=0$. The metric is simplified and takes the form
\begin{equation}
\label{a17}
ds^2=\left(\frac{1-\mu}{1+\mu}\right)^2dt^2-(1+\mu)^4a^2(t)\left[d\tilde{r}^2+\tilde{r}^2\left(d\Theta^2+\sin^2\Theta d\varphi^2\right)\right],\quad \mu=\frac{m}{2\tilde{r}a(t)}.
\end{equation}
Its properties are described in detail in \cite{p18}. For $a$=const (\ref{a17}) goes over to the Schwarzschild metric using the transformation $r=a\tilde{r}(1+\mu)^2=m(2+\mu+\mu^{-1})/2$. The BH horizon corresponds to $\mu=1$ and $r=2m$.

As can be seen, solution (\ref{a16}) goes into both the FLRW metric and the Schwarzschild solution and does not imply modification of either GR or electrodynamics. What prevents us from considering it a solution describing a BH against the background of a homogeneous isotropic cosmological model? It is easy to find the energy-momentum tensor of matter, which should be on the right-hand side of equations (\ref{a1}). It corresponds to an ideal fluid with density $\rho$ and pressure $p$. For solution (\ref{a17}), they can be expressed through the Hubble parameter $H=d(\ln a)/dt$ and its derivative
\begin{equation}
\label{a18}
\rho=\frac{3}{8\pi}H^2,
\end{equation}
\begin{equation}
\label{a19}
p=-\frac{3}{8\pi}H^2-\frac{1}{4\pi}\frac{1+\mu}{1-\mu}\frac{dH}{dt}.
\end{equation}
The density of the fluid depends only on time and is constant in the entire space at a certain moment. There is no concentration or rarefaction of it near BH. The situation with pressure is different. It is the sum of two components. The first is constant at a certain moment at any point in space; the second depends on the coordinate $r$ because $\mu$ depends on it. The second term disappears at $H$=const, when $a$ increases exponentially. In this case, the fluid has an energy-momentum tensor as in the presence of a cosmological constant and the metric (\ref{a17}) turns into the Schwarzschild-de Sitter a.k.a. Kottler metric \cite{p9}, described by (\ref{a4},\ref{a5}) at $Q=0$.

However, if the Hubble parameter changes in time, then the second term in (\ref{a19}) diverges at $\mu=1$. Instead of the BH horizon, the metric describes a weak naked singularity. This is described in more detail in \cite{p18}. Note that when approaching NS towards the center along the radius, the coordinate r is always spacelike. The coordinate t is timelike, but becomes a null vector at $r=2m$.

The problems that prevent the McVittie solution from being considered an exact or nearly exact solution near real BHs are because of equations (\ref{a18},\ref{a19}) in combination with the fact that the Universe contains not only DE, but also matter, both baryonic and dark. The density of matter decreases with time and with it the value of the Hubble parameter. Let me remind you that the accelerated expansion of the Universe means that the value 
\begin{equation}
\label{a20}
\frac{d^2a}{dt^2}=a\left(\frac{dH}{dt}+H^2\right)
\end{equation}
is positive, but the time derivative of $H$ is negative. Therefore, the second term in (\ref{a19}) is positive and increases to infinity as NS is approached. The pressure gradient creates a force that balances the gravitational attraction. Matter does not fall onto McVittie BH. Note also that the equation of state of matter is not barotropic, since pressure of matter varies at different points for the same density. But this is not an incompressible medium, since its density decreases with time along with $H$.

Thus, the McVittie metric cannot describe the objects observed by astronomers with their accretion of matter and its fall onto the BH. However, it can be used to solve a number of problems, e.g., the study of space-time outside a spherically symmetric non-rotating body, but only far from it. At the same time, from (\ref{a18},\ref{a19}) it is clear that it is preferable to use Kottler metric in this case. Studies in which the region near the NS is important may not be entirely adequate for describing astronomical objects. These include discussions of the change in the size of BHs over time or the study of BH stability using quasinormal modes.

\section{Conclusions}

We have considered the problems that arise in various models of black holes, limiting ourselves to the simplest case of spherically symmetric objects. The standard Reissner-Nordstr{\"o}m-de Sitter (RNdS) metric (\ref{a4},\ref{a5}) is an exact solution of the Einstein-Maxwell equations. It is commonly used to describe spacetime near objects discovered by astronomers, if they are not rotating.

McVittie BH \cite{p3} is described by the equations of general relativity and classical electrodynamics, but requires the presence of an exotic medium. Its pressure increases to infinity when approaching the value of the radial coordinate corresponding to the horizon in the Schwarschild solution. Therefore, the object described by this solution is not a BH, but a naked singularity. Nevertheless, the McVittie metric can be used for the study of space-time outside a spherically symmetric non-rotating body, but only far from NS.

Bardeen \cite{p11} and Bardeen-Kiselev \cite{p2} BHs demonstrate a much higher degree of speculation in the model. They were proposed as models of regular black holes, which do not have a central singularity at $r=0$. Therefore, they require the presence of a medium in which the conditions of Penrose's theorem \cite{p10} are violated, for example with a negative energy density. The required very specific form of the energy-momentum tensor of this medium can be introduced in various ways. One can simply postulate its presence as a result of quantum effects, e.g. vacuum polarization, the presence of unknown fields or exotic DE and dark matter, or other currently unknown factors.

The most common explanation involves a combination of several assumptions, each of which is incompatible with the current physical description of the world around us. The black hole has a magnetic charge, and the electromagnetism is described not by the experimentally proven Maxwell equations, but by non-standard electrodynamics, the action of which does not transform into a classical expression in any limit. This action directly includes parameters related to the parameters of a specific BH, i.e. cannot be universal. The anisotropic fluid added by Kiselev must also have unusual properties. It has been shown in Section \ref{S2} that a region of gravitational repulsion may exist around the horizon of a Bardeen-Kiselev BH. It seems to me that all this excludes Bardeen-Kiselev BHs from the list of possible models capable of describing astronomical objects.

In addition to the models considered in the article, there are dozens of other alternative models. Among them, there is not a single one that fully satisfies the requirements of astrophysicists. If we limit ourselves to the case of non-rotating black holes, then the closest to such an ideal model is the study of the behavior of matter against the background of the RNdS metric (\ref{a4},\ref{a5}).

The paper describes a simple method for constructing new models of non-rotating BHs by using a metric of the form (\ref{a4}) with an assumption for a new expression for function $f(r)$. However, the value of any research within such models cannot be compared with the value of research using the RNdS model with function (\ref{a5}), if their goal is to describe the properties of the world around us. Unfortunately, articles based on alternative models often hide this fact. Their conclusions are sometimes formulated in such a way as to claim universality and description of the properties of astronomical objects. Let me clarify that I am not suggesting that we stop using other models, but rather that studies based on different models vary greatly in their value for describing real massive compact objects in space, currently identified as BHs.\\

The following abbreviations are used in this preprint:\\

\noindent 
\begin{tabular}{@{}ll}
BH & Black hole\\
DE & Dark energy\\
FLRW & Friedmann-Lema\^{i}tre-Robertson-Walker\\
GR & General Relativity\\
NS & Naked singularity\\
RNdS & Reissner-Nordstr{\"o}m-de Sitter\\
\end{tabular}


\end{document}